 \documentclass[prb,preprint,amsmath,amssymb]{revtex4}
\usepackage{color}
\begin{document}
\author{Kevin Leung$^{*}$, Louise J.~Criscenti, Andrew W.~Knight,
Anastasia G.~Ilgen, Tuan~A.~Ho, and Jeffery A.~Greathouse}
\affiliation{Sandia National Laboratories, MS 1415, \& 0754,
Albuquerque, NM 87185\\
$^*${\tt kleung@sandia.gov}}
\date{\today}
\title{Concerted Metal Cation Desorption and Proton Transfer
on Deprotonated Silica Surfaces} 

\input epsf

\begin{abstract}

The adsorption equilibrium constants of monovalent and divalent cations 
to material surfaces in aqueous media 
are central to many technological, natural, and geochemical processes.  
Cation adsorption/desorption is often proposed to occur in concert with
proton-transfer on hydroxyl-covered mineral surfaces, but so far this
cooperative effect has been inferred indirectly.
This work applies Density Functional Theory (DFT)-based molecular dynamics
simulations of explicit liquid water/mineral interfaces to calculate metal
ion desorption free energies.  Monodentate adsorption of Na$^+$, Mg$^{2+}$,
and Cu$^{2+}$ on partially deprotonated silica surfaces are considered.
Na$^+$ is predicted to be unbound, while Cu$^{2+}$ exhibits larger binding
free energies to surface SiO$^-$ groups than Mg$^{2+}$.  The predicted trends
agree with competitive adsorption measurements on fumed silica surfaces.  
As desorption proceeds, Cu$^{2+}$ dissociates one of the H$_2$O molecules in
its first solvation shell, turning into Cu$^{2+}$(OH$^-$)(H$_2$O)$_3$, while Mg
remains Mg$^{2+}$(H$_2$O)$_6$.  The protonation state of the SiO$^-$ group at
the initial binding site does not vary monotonically with cation desorption.  

\end{abstract}

\maketitle

The adsorption free energies of ions on to material surfaces in liquid
media govern equilibrium constants and adsorption isotherms, which are
central to many technological and geochemical processes.\cite{gibbs2017,rosso2018,peng2014,hayes1987,ni2014}
One key aspect of interfaces between aqueous electrolyte and materials
interface, distinct from those of aprotic solvents, is the
possibility of acid-base reactions occurring in conjunction with ion
adsorption.  Surface hydroxyl (-OH) groups, ubiquitous at water interfaces,
can release H$^+$ as cations adsorbs, especially for multivalent cations.
OH$^-$ can also be released as oxyanions bind to the surface.\cite{hayes1987}
This phenomenon is well known in geochemistry
contexts,\cite{sverjensky2006,pbse,geiger2010} but its occurrence is generally
inferred indirectly from measurements or estimated from continuum models.
Computationally, it has been shown that divalent
cation binding energies at water-oxide interface can differ by a very large
amount -- $\sim 2$~eV (46~kcal/mol)\cite{kerisit2015} -- depending on whether
surface OH groups are deprotonated or not.  But the computational method used
there, and in most existing molecular calculations, does not permit spontaneous,
simultaneous deprotonation and cation adsorption events.  It begs the question
of what quantitative effect allowing such cooperative behavior would make.
Benchmark calculations are urgently needed.

Apart from this cooperativity, understanding many aspects of cation adsorption
remains challenging.  Tremendous progress has been made via
non-linear optical spectroscopy,\cite{geiger2010} nuclear magnetic
nesonance (NMR),\cite{nmr} X-ray photoelectron spectroscopy and other X-ray
methods,\cite{hayes1987,cheah2000,cheah1998}
titration,\cite{cheah1997} batch adsorption,\cite{elliot1986,subramaniam}
and calorimetry techniques\cite{kabengi2017} to investigate cation adsorption,
including on silica surfaces which are ubiquitous in many applications and areas
of science.  But disagreements about fundamental aspects, such as whether
cations are adsorbed in outer-\cite{geiger2010,nmr} or
inner-sphere\cite{kabengi2017} configurations, have persisted.  Here
inner-sphere means direct contact between SiO$^-$/M$^+$, while outer-sphere
is water-mediated adsorption, SiO$^-$/H$_2$O/M$^+$.

Atomic length-scale modeling can shed light on many aspects of cation
adsorption.\cite{hocine2016,criscenti2013,kerisit2015,hartkamp2015,kalinichev}
In particular, force field-based potential-of-mean-force (PMF) calculations
have been successfully applied to study divalent metal cations desorption from
mineral surfaces.\cite{kerisit2015,criscenti2013}
However, it is challenging to use non-electronic structure methods to model
the interactions between transition metal (TM) cations and ligands.\cite{pbse}
This is because TM ions with partially filled $3d$ orbitals are difficult to
parameterize with molecular force fields.  Furthermore, with few
exceptions,\cite{vanduin,garofalini,rimsza2018} most existing force fields for
minerals do not permit dynamical changes in the protonation states in the
ubiquitous hydroxyl groups on mineral surfaces.  Molecular dynamics studies
have generally resorted to static assignment of surface hydroxyl deprotonation
states.  This approach cannot address correlations and dynamical fluctuations
in the electrostatic environments of binding sites due to acid-base reactions.

The present work highlights the coupled cation desorption/material surface
proton transfer  phenomenon, using Density Functional Theory (DFT)-based
molecular dynamics (or ``{\it ab initio} molecular dynamics''/AIMD) simulations.
AIMD has been shown to predict acid-base reactions to within 1~pH
unit.\cite{liu13a,sulpizi2014,pfeiffer16,pfeiffer16a,pfeiffer16b,pfeiffer15,sulpizi12,sulpizi12a,cheng1_sprik,leung2009,leung2013,kubicki}
DFT also explicitly includes the effect of partial $3d$ orbital occupancies.
AIMD is therefore well-suited to modeling metal adsorption on mineral surfaces
with OH groups at binding sites.\cite{pfeiffer16b,pbse,meijer2016,meijer2017}  
Existing AIMD work have not directly addressed cooperative acid-base/cation
adsorption, partly because some materials surfaces previously
examined may not exhibit this behavior and partly because PMF methods have
not been applied to pull cations sufficiently far from the surface.
Our AIMD results will be supported by batch adsorption isotherm experiments
and compared with classical force field predictions.

The cations of interest here are Na$^+$, Mg$^{2+}$, and Cu$^{2+}$.
The model substrate chosen is a reconstructed $\beta$-crystabolite (001)
surface, with about 4/nm$^2$ SiOH group surface density.\cite{leung2009}
It has several advantages as a benchmark.  (1) The surface SiOH density
is in reasonable accord with that cited for well-soaked, amorphous
silica.\cite{brinker} (2) We have previously computed its
pK$_a$ using AIMD potential-of-mean-force calculations, and can therefore
correlate this pK$_a$ with cation desorption calculations conducted using
similar AIMD methods.  (3) The surface has a moderate cell size, and only
one distinct type of silanol (SiOH) group.  \color{black} Each SiOH is
at least 5~\AA\, from all other SiOH, ensuring that cations can only coordinate
to one SiO$^-$ group on this model surface (i.e., they are monodentate).  
\color{black}  This feature simplifies the computational analysis.  
In contrast, many crystalline mineral surfaces feature numerous cation-binding
sites and types of hydroxyl groups.\cite{pbse} 

Finite temperature AIMD simulations apply the
Perdew-Burke-Ernzerhof (PBE) functional,\cite{pbe} the projector-augmented
wave-based Vienna Atomic Simulation Package (VASP),\cite{vasp,vasp1} a
400~eV energy cutoff, and $\Gamma$-point sampling of the Brillouin zone.
All simulation cells have dimensions 10.12\AA$\times$10.12\AA$\times$26.0~\AA. 
These settings are similar to those in our previous pK$_a$
work.\cite{leung2009,leung2013} Spin-polarized DFT calculations, with one
net unpaired electron, are conducted when Cu$^{2+}$ is present.  The Cu
pseudopotential used does not include pseudovalent $3p$ electrons.  
To keep the simulation cells charge neutral, one or two H$^+$ are removed
from the silica surface with an adsorbed cation.  One of the resulting
SiO$^-$ is initially coordinated to the M$^{n+}$.  The overall stoichiometries
are Si$_{20}$O$_{107}$H$_{133}$$^-$Na$^+$ and
Si$_{20}$O$_{108}$H$_{134}^-$M$^{2+}$.  For details of
the initiation of AIMD simulation cells, see the Supporting Information 
(S.I.) document.

In the presence of acid functional groups at water/material interfaces, the
pH in the simulation cell should be pinned at the pK$_a$ of functional
groups, provided that (1) there is only one type of such groups; (2) a
fraction of them are deprotonated; (3) their pK$_a$ is lower than that
of H$_2$O; and (4) the surface groups do not interact with each other.
Within the non-interacting assumption, the pH in our AIMD cells should be
between 7.0~and 8.1 -- the pK$_a$ range previously predicted for this
surface.\cite{leung2009} In experimental samples with amorphous or crystalline
silica, bimodal or trimodal pK$_a$ distributions of pK$_a$ have been
reported.\cite{shen,julie}  It would have been more challenging to
assign pK$_a$ in AIMD simulation cells with multiple types of SiOH.

Potential-of-mean-force (PMF) free energy simulations computes $\Delta W(Z)$
as the natural logrithm of the probability distribution of the two-body
reaction coordinate of the form $Z$=$z_{\rm O}-z_{\rm M}$ (S.I.).
Here $Z$ is the coordinate normal to the silica-water interface, M is the
desorbing cation, and O is part of the initially deprotonated surface SiOH
group bound to M$^{n+}$.  The first window represents configurations with
M$^{n+}$ bound to the SiO$^-$ group; subsequent windows have M$^{n+}$
progressively displaced farther away.

$\Delta W(Z)$ is effectively the constrained free energy at a $Z$ value; it
includes neither fluctuations around $Z$ nor the standard state reference
associated with aqueous solutions.  To obtain the adsorption free energy
($\Delta G_{\rm ads}$) from $\Delta W(Z)$, we integrate configuration space in
three dimensions, and account for  the entropic contribution from a standard
state 1.0~M ideal concentration solution:\cite{klein} 
\begin{equation}
\Delta G_{\rm ads} /k_{\rm B} T = -\log \{  \int_\Omega d\Omega 
		\exp [-\Delta W(Z)/k_{\rm B}T]  /(V_o)\} \, . \label{eq1}
\end{equation}
Here $V_o$ is the volume associated with 1.0~M aqueous solution (1662~\AA$^3$)
and T=300~K is assumed.  
The volume element $\Omega$ spans the configuration space where M$^{n+}$
is ``bonded'' to the SiO$^-$ group.  A limiting bonding distance of 
2.50~\AA\, is assumed.  At this separation the pair correlation functions
between transition metal ions and water oxygen sites exhibits their first
turning points.\cite{pasq2001}  
The angular distribution is also involved in the
integral.  To our knowledge, $\Omega$ has not been standardized
for PMF calculations at interfaces.\cite{criscenti2013,meijer2017,kerisit2015}
Here we approximate it as a cylinder with a radius $R$=0.5~\AA\, estimated
from a completely unconstrained trajectory.  Electrostatic corrections
associated with image dipoles are added to Mg$^{2+}$ PMF predictions.  These
corrections do not exceed 0.1~eV, are less than 0.03~eV for monovalent cations,
and are further discussed in the S.I.

\begin{figure}
\centerline{\hbox{ \epsfxsize=4.00in \epsfbox{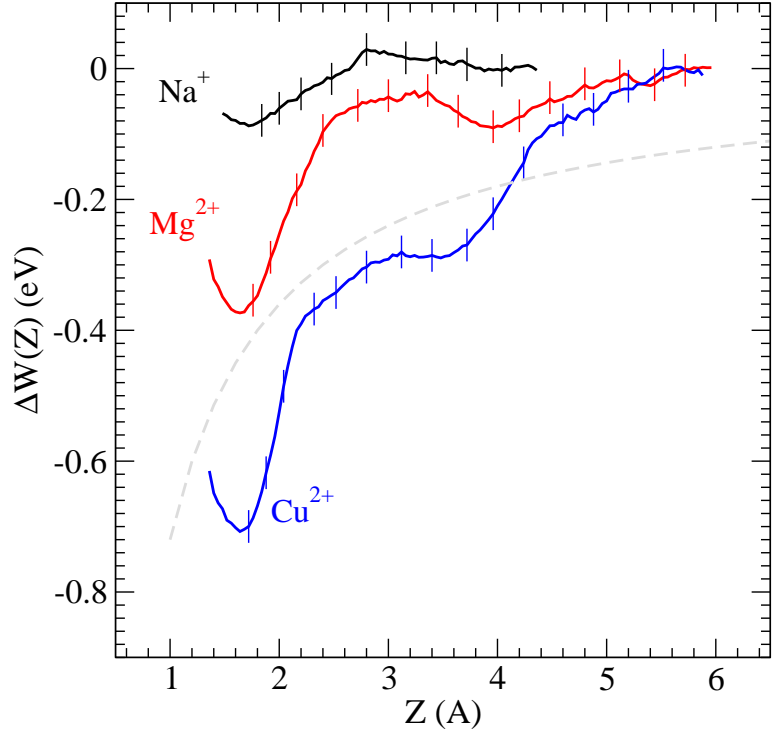} }}
\caption[]
{\label{fig1} \noindent
Potential-of-mean-force (PMF) predictions of metal cation M$^{n+}$
desorption profile ($\Delta W(Z)$) as a function of the vertical
separation ($Z$) between Mn$^{n+}$ and the O atom of the SiO$^-$ group to
which the cation is initially coordinated.  Black, red, blue refer to
M=Na, Mg, and Cu, respectively.  Grey dashed line denotes the coulomb attaction
between +2 and -2 charges screened by $\epsilon_o$=78.  Vertical lines indicate
the end points of sampling windows; they are not error bars.
}
\end{figure}

Figure~\ref{fig1} depicts the potentials-of-mean-force associated with
Na$^+$, Mg$^{2+}$, and Cu$^{2+}$ desorption.  The minima of $\Delta W(Z)$,
relative to the asymptotic, large $Z$ regime, are -0.087$\pm$0.03~eV,
-0.38$\pm$0.05~eV, and -0.71$\pm$0.07, respectively.  The uncertainties
reflect one standard deviation.

First we focus on Na$^+$ on reconstructed $\beta$-cristobalite surface, with
one deprotonated SiOH group in the surface cell. In other words, 25\% of
the SiOH on one surface is deprotonated.  Fig.~\ref{fig1} shows that the
$\Delta W(Z)$ minimum is very shallow (-0.087~eV) for Na$^+$.  
We start an AIMD trajectory with Na$^+$
coordinated to the surface site, without applying constraints.  Na$^+$ remains
there for $\sim$45~ps; at longer times, it spontaneously escapes into the
bulk solution.  The statistics in Fig.~\ref{fig1} are collected prior to
this spontaneous event.  In subsequent sampling windows,
PMF calculations circumvents this time scale problem by using constraining
potentials.  Substituting $\Delta W(Z)$ into Eq.~\ref{eq1}, the free energy
cost of Na$^+$ adsorbing on to this silica surface from a 1.0~M Na$^+$
solution is $\Delta G_{\rm abs}$=+0.13~eV, if Na$^+$ is assumed to exhibit
an activity coefficient of unity.  The positive sign of $\Delta G_{\rm abs}$
means there is no tendency for Na$^+$ to bind to the surface.  

\begin{figure}
\centerline{\hbox{ \epsfxsize=1.40in \epsfbox{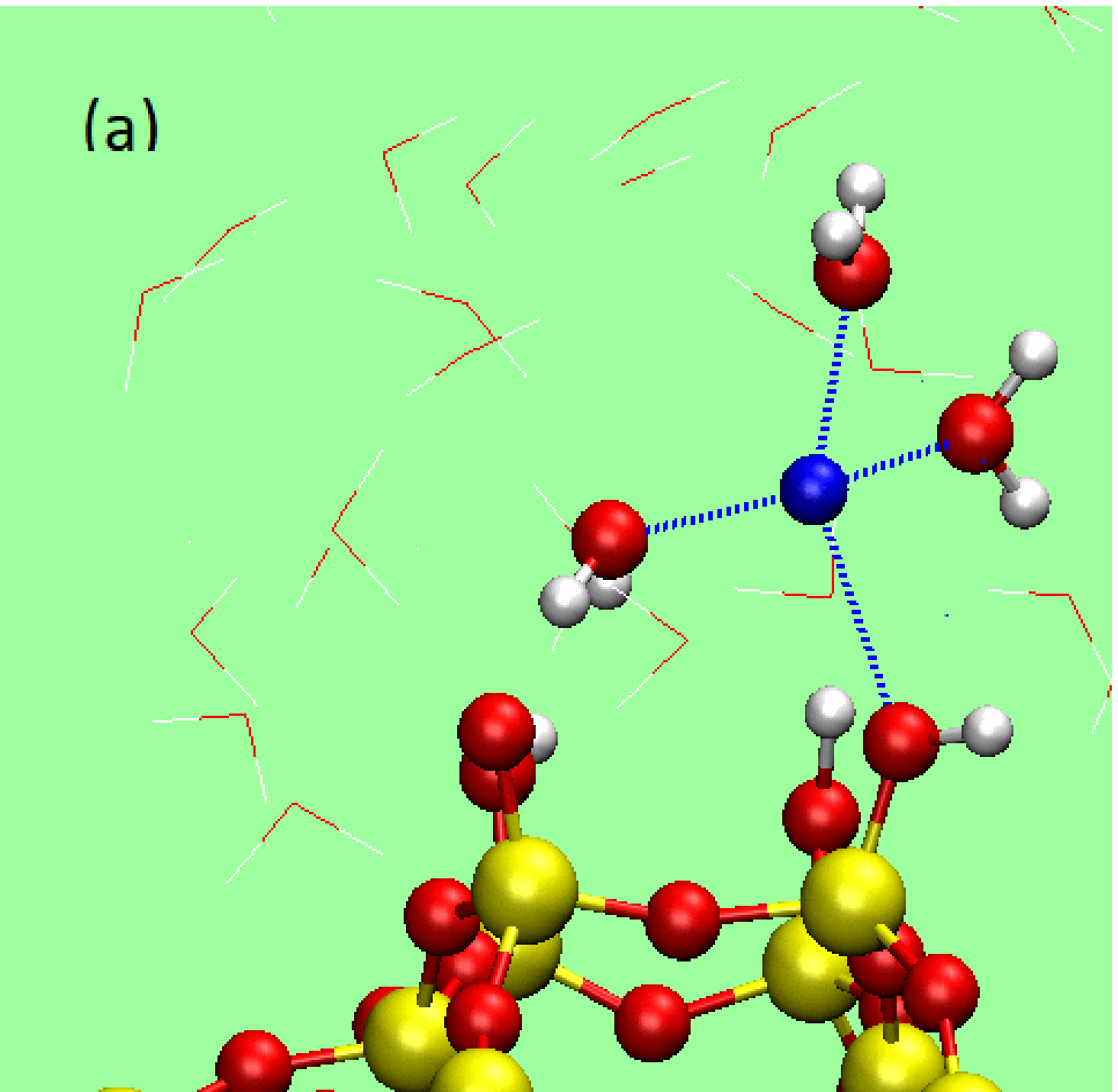} 
		   \epsfxsize=1.40in \epsfbox{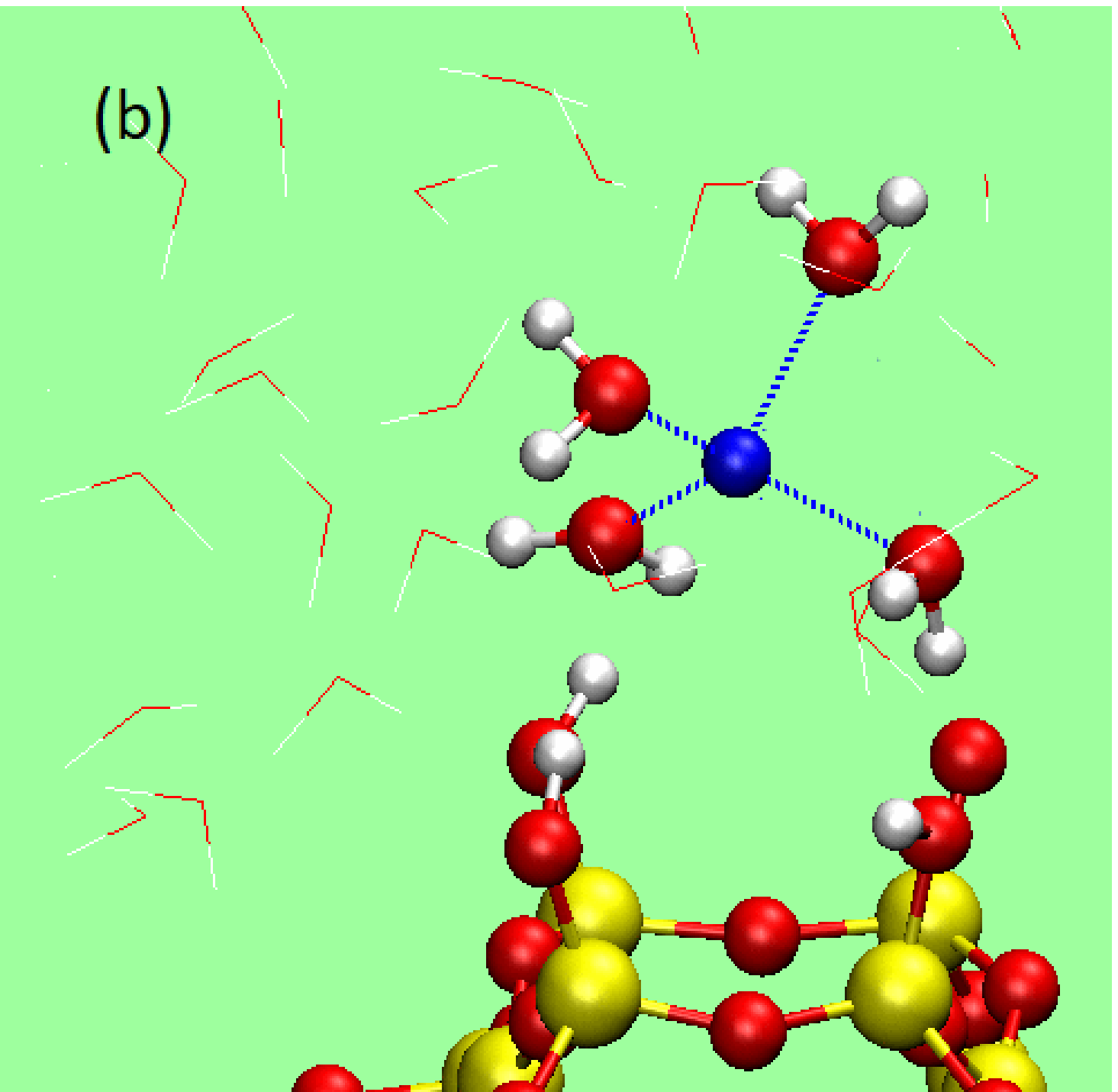} 
		   \epsfxsize=1.40in \epsfbox{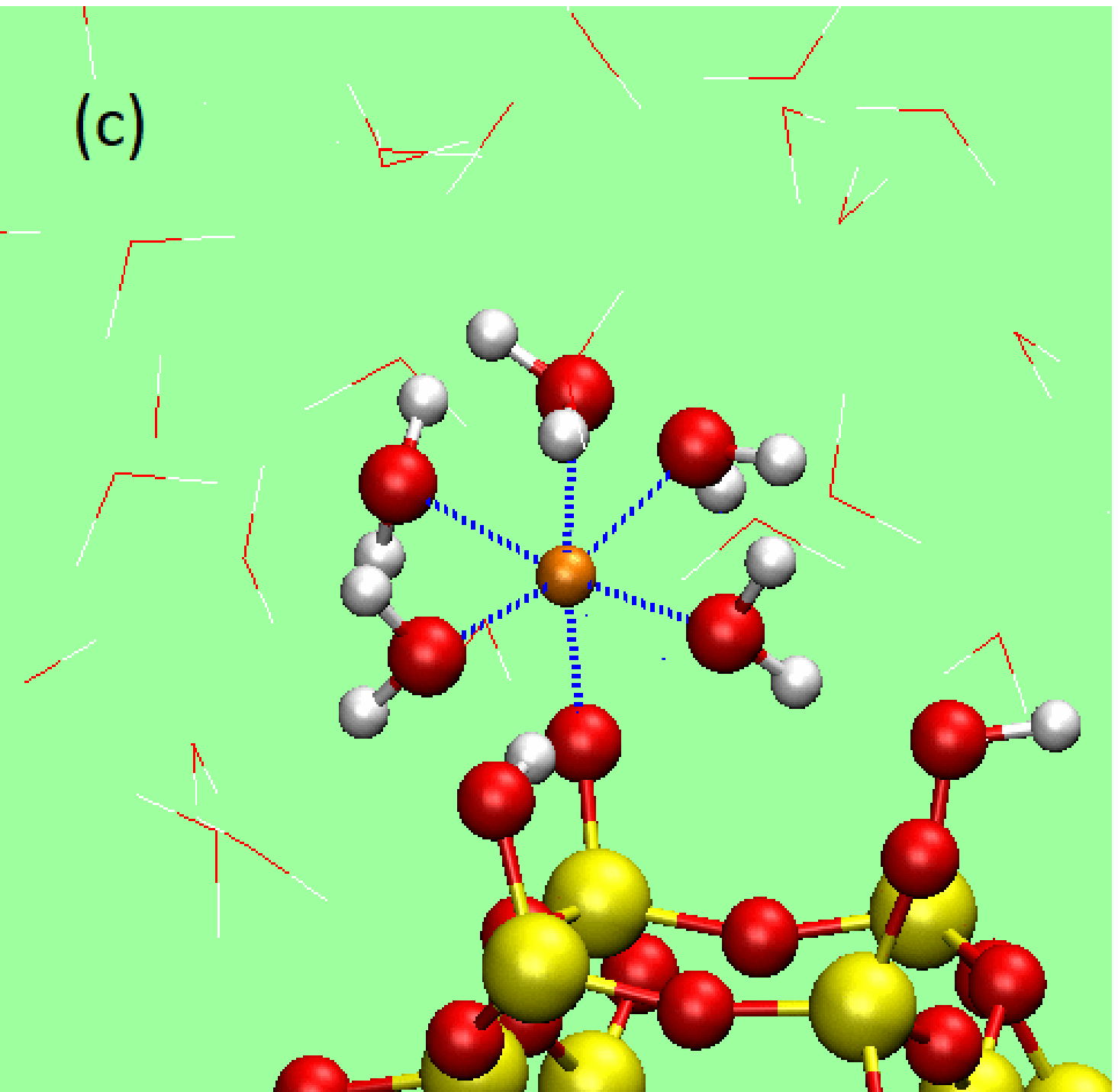} 
		   \epsfxsize=1.40in \epsfbox{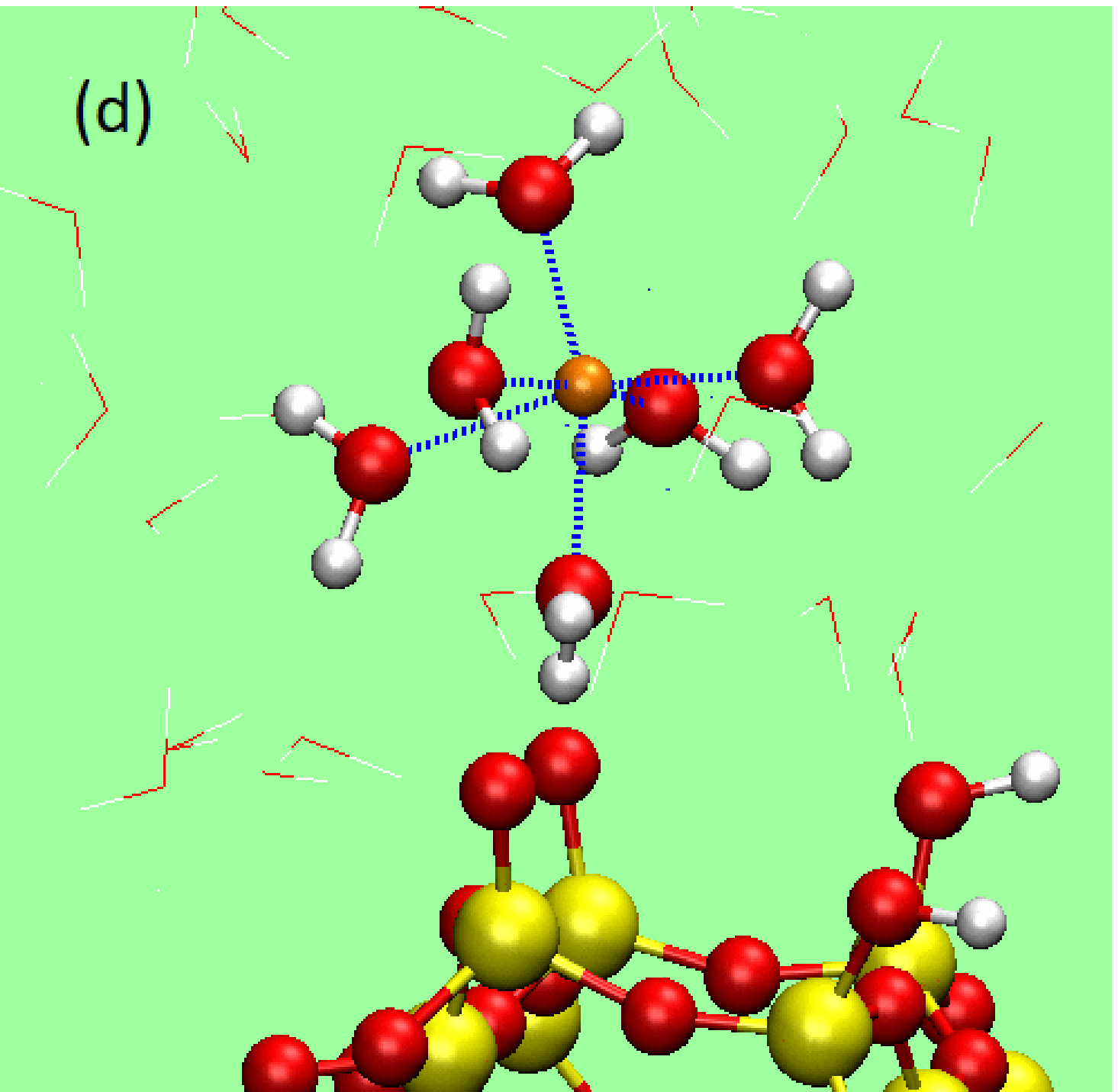} }}
\centerline{\hbox{ \epsfxsize=1.40in \epsfbox{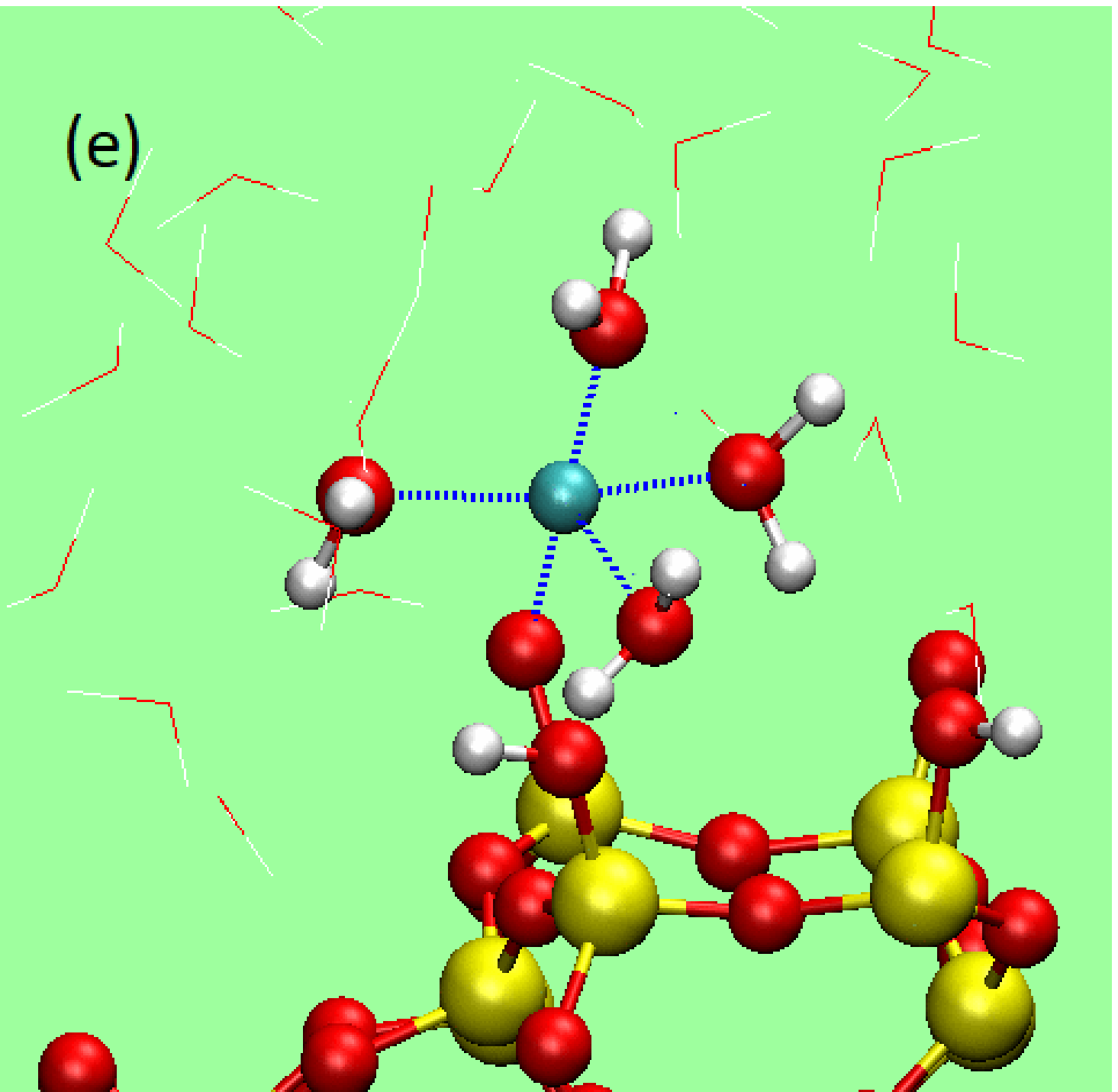} 
		   \epsfxsize=1.40in \epsfbox{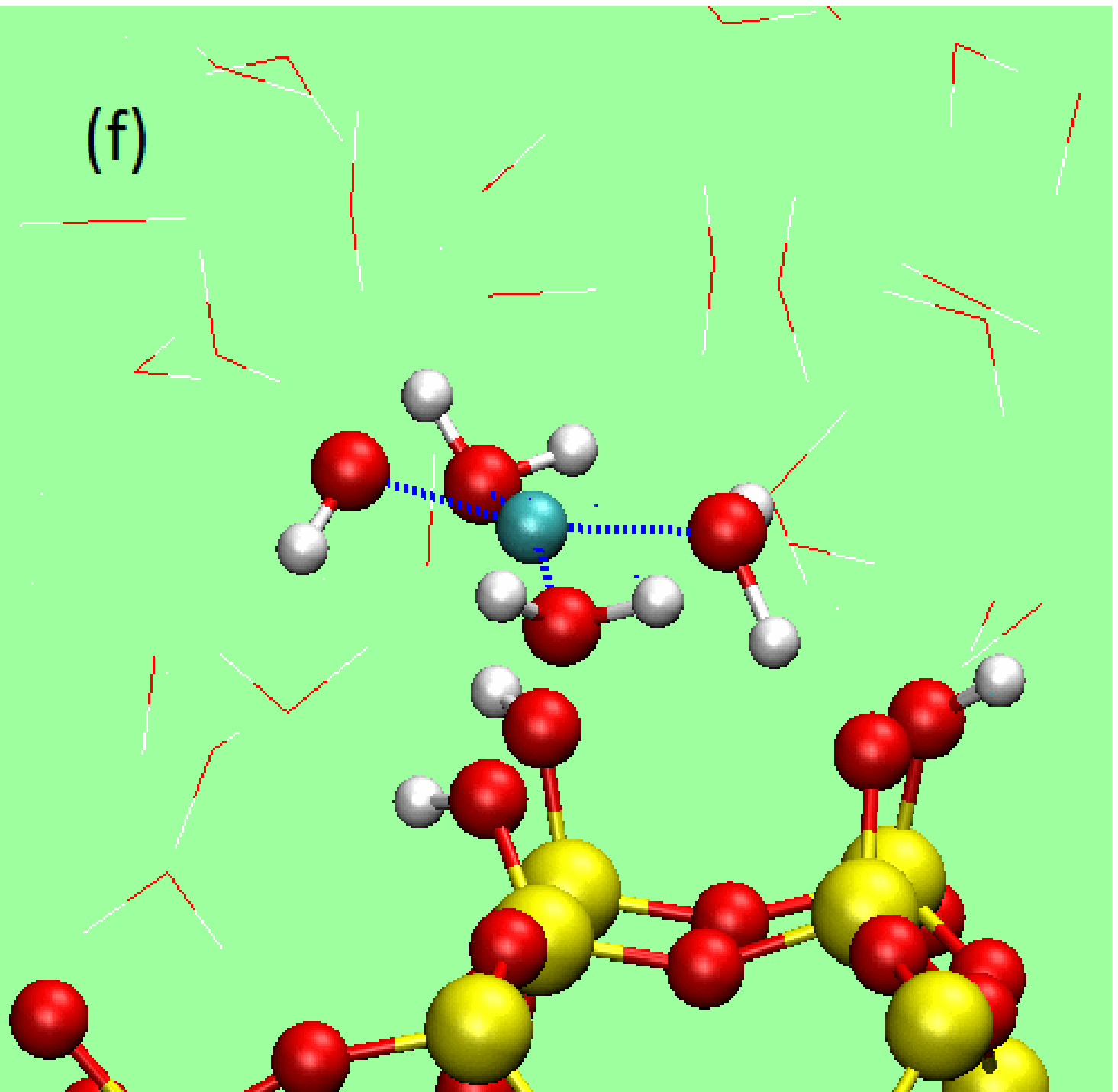} 
		   \epsfxsize=1.40in \epsfbox{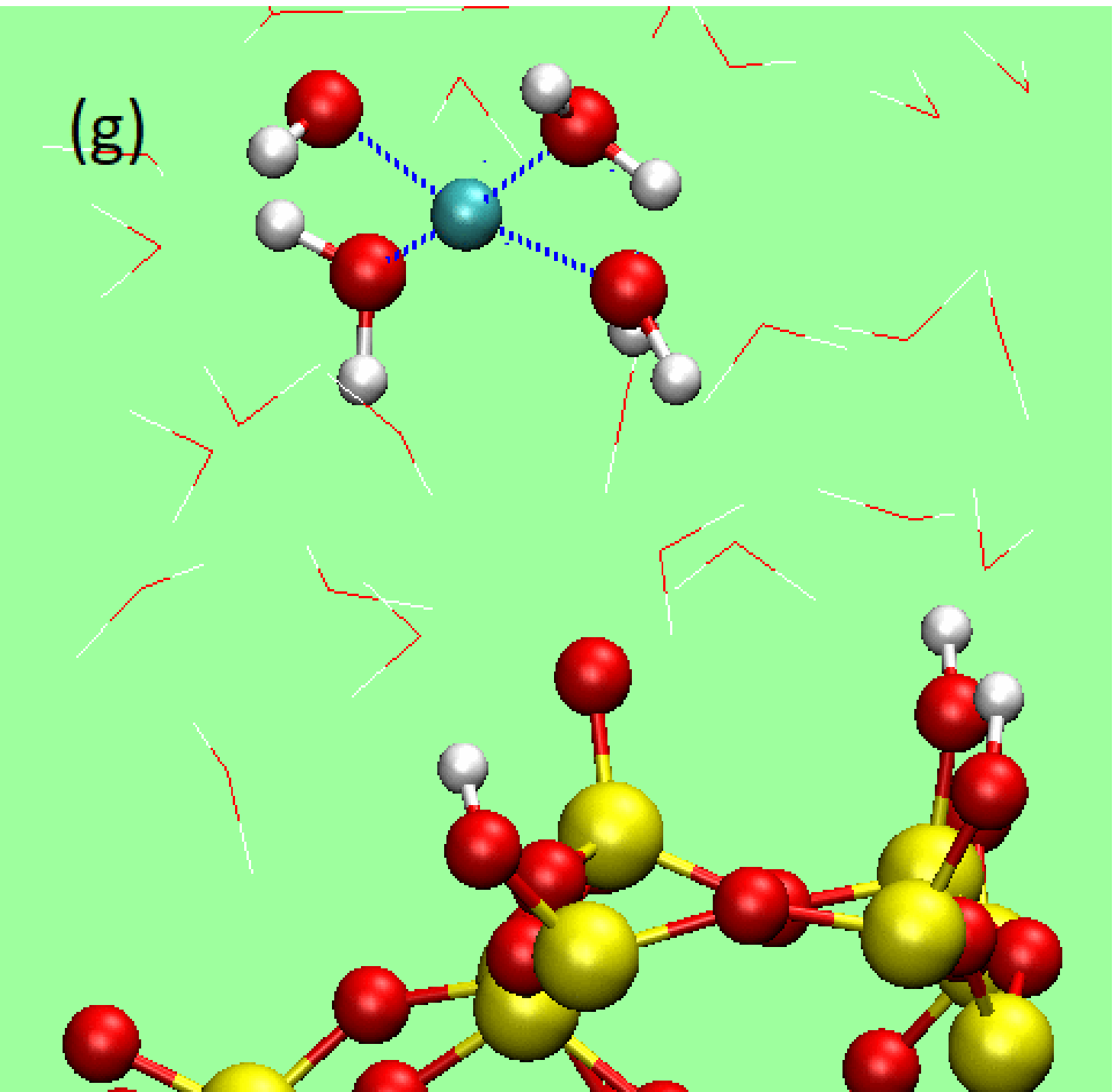} }}
\caption[]
{\label{fig2} \noindent
Snapshots of (a)-(b) Na$^+$, $Z$$\sim$2.7~\AA, and $3.0$~\AA;
(c) Mg$^{2+}$, $Z$$\sim$1.6~\AA\, (unconstrained in $Z$ direction);
(d) Mg$^{2+}$, $Z$$\sim$3.8~\AA;
(e) Cu$^{2+}$, $Z$$\sim$1.6~\AA;
(f) Cu$^{2+}$, $Z$$\sim$2.7~\AA;
(g) Cu$^{2+}$, $Z$$\sim$4.5~\AA.
Na, Mg, \color{black} Cu, \color{black}
Si, O, and H atoms are colored in blue, orange, cyan, yellow,
red, and white.  H$_2$O are thin lines, except that those coordinated to
the cation are depicted as ball-and-stick models.  Thicker lines depict
SiO-H bonds. No OH$^-$ is coordinated to the cations in panels (a)-(e), but
a OH$^-$ clearly appears on Cu$^{2+}$ in (f)-(g).
}
\end{figure}

Fig.~\ref{fig2}a-b depict configurations in trajectories constrained
around $Z$$\sim$2.7~\AA\, and 3.0~\AA, respectively.  In both snapshots,
the SiO$^-$ group originally coordinated to Na$^+$ has become protonated.
An H$_2$O molecule has spontaneously inserted between the SiOH group and
the Na$^+$ before the second snapshot takes place.  \color{black} An apparent
coordination number of four is seen in these snapshots only because
the Na$^+$ is close to the surface.  In the $Z$$\sim$4.2~\AA\, window, 
we have computed the pair correlation $g(r)$ between Na$^+$ and H$_2$O
O~atoms.  The first minimum in the $g(r)$ is $\sim$3.16~\AA; integrating
to this distance yields an average hydration number of 5.17.  These
values are similar to those computed for Na$^+$ in bulk water using
AIMD\cite{rowley} and polarizable force field\cite{ponder} methods.\color{black}

A recent calorimetry study reports favorable inner-sphere Na$^+$ adsorption
enthalpy on partially deprotonated quartz particle surfaces at
pH=4.\cite{kabengi2017}  Relating the quantitative enthalpic value per
milligram silica sample to our calculations is challenging because the
surface densities per type of deprotonated SiOH groups per unit surface area
are not specified.  Na$^+$ adsorption on silica has been inferred as
inner-sphere\cite{kabengi2017} or outer-sphere,\cite{nmr} depending on the
measurement technique used, although the details of the silica samples and
electrolytes differ.  Our prediction that Na$^+$ is unbound on a specific
model silica surface beyond a $\sim$45~ps residence time may help interpret
the data and resolve the discrepancy.  

Recent AIMD simulations of deprotonated silica
surfaces\cite{pfeiffer16a,kubicki} have reported persistent Na$^+$ coordination
to the quartz (001) surface for the duration of trajectory lengths of
$\sim$10~ps.  However, that surface exhibits a higher SiOH surface density
than our model or amorphous silica. Na$^+$ can be also be bidentate or
even tridentate on quartz (001).  
Future AIMD calculations to compute desorption PMF associated
with Na$^+$ at bidentate sites will further clarify transient inner- versus
outer-sphere adsorption behavior.

We have also performed classical MD simulations using the
same AIMD cristobalite surface, but increasing the surface area by a factor of
four and adding a thicker aqueous layer (30~\AA). Adsorption properties
are obtained from unconstrained and umbrella sampling simulations.  Final
snapshots reveal that the Na$^+$ moves away from the SiO$^-$ site and
forms an inner-sphere monodentate complex with a neutral SiOH site.  This
is qualitatively similar to ReaxFF predictions.\cite{rimsza2018}  Although
the classical MD Na$^+$/SiO$^-$ interactions have not yet been fitted to
DFT/AIMD predictions and differ from AIMD simulations in the binding site
configuration, the classical $\Delta W(Z)$ is qualitatively similar (Fig.~S3
in the S.I.).  

The first minimum in the Mg$^{2+}$ $\Delta W(Z)$, at -0.37~eV, is
located at $\sim$1.68~\AA\, (Fig.~\ref{fig1}).  A weak second minimum 
at 3.96~\AA\, yields a -0.09~eV attraction, separated by a small barrier
at 3.36~\AA.  This second valley represents an outer-sphere complex.
In all configurations examined, Mg$^{2+}$ is octahedrally
coordinated to either 5 H$_2$O molecules and one SiO$^-$ group
(Fig.~\ref{fig2}c) or 6 H$_2$O molecules (Fig.~\ref{fig2}d).  
Integrating $\Delta W(Z)$ via Eq.~\ref{eq1} yields a
favorable 0.14~eV binding free energy, unlike Na$^+$ which is predicted
to be unbound.  Calorimetry studies has shown that Mg$^{2+}$ adsorption 
yields larger enthapy release than Na$^{+}$.\cite{kabengi2017}
Reference~\onlinecite{geiger2010} has estimated a small 0.30~eV
binding free energy between Mg$^{2+}$ and silica surfaces, although it
interprets the coordination as outer-sphere.


The grey dashed line depicts the coulombic attraction between $\pm$2 charges
screened by 1/$\epsilon_o$ appropriate to liquid water.  The $\Delta W(Z)$
curve for Mg$^{2+}$ does not exhibit such asymptotic behavior.  Other
classical force field-based PMF calculations with charged mineral surfaces
and larger simulation cells than AIMD simulations also reach a plateau
for $Z$$\sim$6~\AA; see Ref.~\onlinecite{kerisit2015} and Fig.~S3 in this
work.  Hence the Mg$^{2+}$ plateau in Fig.~\ref{fig1} does not appear related
to finite size effects associated with AIMD simulation cells.

Fig.~\ref{fig1} also depicts the $\Delta W(Z)$ associated with Cu$^{2+}$.
Substituting the relevant $\Delta W(Z)$ into Eq.\ref{eq1}, the Cu$^{2+}$
adsorption free energy is found to be a favorable -0.47~eV.  This is 
substantially higher than our predicted Mg$^{2+}$ value.  It is a
general trend that first row transition metal divalent cations tend to 
exhibit stronger binding behavior.\cite{langmuir}
In the SiO$^-$-Cu$^{2+}$ inner-sphere complex regime (Fig.~\ref{fig2}e),
Cu$^{2+}$ is coordinated to 4~H$_2$O and a SiO$^-$ group, which is slightly
different from the 6-coordinated Mg$^{2+}$ (Fig.~\ref{fig2}d).  
Indeed, even in liquid water, it has been reported that the inner solvation
shell of Cu$^{2+}$ exhibits first hydration shell behaviors distinct from
other first row transition metal ions.\cite{pasq2001,cheah2000}  

Away from the $Z$$>$2~\AA\, regime, our AIMD simulations consistently predict
that Cu$^{2+}$ exists as a square-planar Cu$^{2+}$(OH$^-$)(H$_2$O)$_3$ instead
of an octahedral Cu$^{2+}$(H$_2$O)$_6$ (Fig.~\ref{fig2}f-g).  The OH$^-$
originates from an H$_2$O molecule directly coordinated to Cu$^{2+}$ which has
lost a H$^+$ to one of the two SiO$^-$ groups initially on the surface when
$Z$ is small.  This H$_2$O deprotonation event changes the Cu$^{2+}$
coordination number drastically.

We have not applied AIMD to calculate the pK$_a$ associated with H$_2$O
moecules bonded to Cu$^{2+}$.  But experimental pK$_a$
of water in the M$^{2+}$(H$_2$O)$_6$ first solvation shell have been quoted
as 8 and 11.2-11.4 for Cu$^{2+}$ and Mg$^{2+}$,
respectively.\cite{stumm1996,baes1976}  Similar trends have been predicted
in cluster-based DFT calculations.\cite{dixon2015} The silanol deprotonation
pK$_a$ for this reconstructed $\beta$-cristobalite surface has been
predicted to be 7.0-8.1.\cite{leung2009}  So the pK$_a$ of
SiOH and Cu$^{2+}$(H$_2$O)$_6$ are very similar.  The fact that the simulation
cell contains a fraction of deprotonated SiOH means that the solution pH
is pinned at the SiOH pK$_a$ value.  It is therefore reasonable that we observe
deprotonation of Cu$^{2+}$(H$_2$O)$_n$ but not of Mg$^{2+}$(H$_2$O)$_6$.  
A lower pK$_a$ in the first hydration shell has been linked to stronger
tendency towards metal cation adsorption or precipitation.\cite{elliot1986}
Experimentally, SiOH pK$_a$ has been proposed to follow bimodal\cite{shen}
or even trimodal\cite{julie} distributions.  If Cu$^{2+}$ is desorbed from
a SiO$^-$ group with pK$_a$=4.5,\cite{shen} the
Cu$^{2+}$(H$_2$O)$_n$ complex would be insufficiently acidic to protonate the
SiO$^-$ group.  So far, at atomic length-scale, such low pK$_a$ SiOH groups
have been identified only on quartz (0001) or amorphous silica
surfaces,\cite{pfeiffer16,pfeiffer16a,pfeiffer16b} or as transient
species.\cite{leung2009}

\color{black}
In addition to the deprotonation of a H$_2$O coordinated to 
Cu$^{2+}$, AIMD simulations of all three cations involve proton exchange
between neighboring SiO$^-$ and SiOH groups on the same surface via 
water bridges once the cations are sufficiently far away from the surface.
This will be further discussed below (Fig.~\ref{fig4}).
\color{black}

In Fig.~S3,
classical force field MD simulations also yield a larger Cu$^{2+}$ desorption
free energy than Mg$^{2+}$.  This is qualitatively consistent with AIMD
results, although the magnitude of the inner-shell minimum in $\Delta W(Z)$
is overestimated compared to AIMD predictions, just as it is for Mg$^{2+}$.
A separate simulation was performed with a hydroxide ion added to the
Cu$^{2+}$ system. The hydroxide ion remains in the first coordination
shell of Cu$^{2+}$, but the Cu$^{2+}$ coordination changes from 6-fold
to 5-fold when the hydroxide ion is present.  The $\Delta W(Z)$ for both
ions are predicted to be qualitatively similar but with larger Cu$^{2+}$
inner-shell solvation free energy.  It cannot be
ruled out that part of the overestimate lies in the inability of classical
force field used to adequately describe cooperative acid-base behavior.
Fine-tuning of the force field will be conducted in the future.

There are few AIMD simulations of Cu$^{2+}$ on oxide surfaces, but predictions
about other transition metal ions make useful comparisons.  
Ref.~\onlinecite{meijer2017} applies AIMD simulations to show that removing
Ni$^{2+}$ from SiOH groups at clay edges requires a significant 0.69~eV to
reach the first barrier at $\sim$3~\AA.  This 0.69~eV free energy change
is likely associated with an outer-sphere configuration rather than infinite
separation between Ni$^{2+}$ and the clay substrate.  Cd$^{2+}$ is predicted
to exhibit a smaller, but still significant 0.23~eV initial barrier towards
desorption.\cite{meijer2016}  Using classical force field-based molecular
dynamics, Ref.~\onlinecite{kerisit2015} models Fe$^{2+}$ desorption from
both charge-neutral and deprotonated hematite (001) surfaces.  While a
quantative comparison with our charged surface should not be made, the
PMF reported there for the neutral surface has less structure in 
$\Delta W(z)$ beyond $z=4$~\AA\, than in the Mg$^{2+}$ PMF in
Ref.~\onlinecite{criscenti2013} regardless of the force field used.  The
charged surface with a (SiO$^-$)$_2$Fe(II) binding site exhibits
$\sim$1.7-2.2~eV first minima in $\Delta W(Z)$.  We stress that none of the
calculations discussed for this comparison purpose have reported cooperative
acid-base reactions -- either because they apply classical force fields or
because the material surface is different.  

On amorphous silica surfaces, Cu$^{2+}$ can form dimers.\cite{cheah1998}
At sufficiently high pH it precipitates to form Cu(OH)$_2$.
Our AIMD simulation cells, which contain one divalent cation, cannot
model dimerization and precipitation.  However, the intriguing desorption
configurations we predict are likely precursors to such complex
Cu$^{2+}$ behavior if more than one Cu$^{2+}$ is present.

\begin{figure}
\centerline{\hbox{ \epsfxsize=4.00in \epsfbox{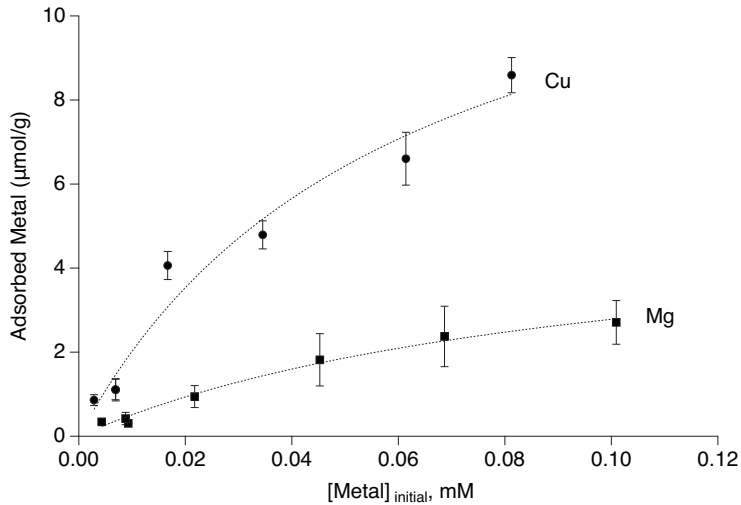} }}
\caption[]
{\label{fig3} \noindent
\color{black} Competitive batch adsorption measurements showing that Cu$^{2+}$
adsorbs more readily than Mg$^{2+}$ on fumed silica surfaces at different
electrolyte concentrations.
\color{black}
}
\end{figure}

Batch adsorption isotherm experiments were performed to evaluate the
competitive adsorption of Cu and Mg on fumed silica and provide
validation of DFT calculations.
For competitive adsorption studies, Cu(NO$_3$)$_2$ and Mg(NO$_3$)$_2$ were
added to each 50 mL centrifuge tube with concentrations ranging from
0.005~mM to 0.1 mM and brought up to 10 mL total volume with Milli-Q water.
The pH was adjusted to 6.0 $\pm$ 0.1 using dilute HNO$_3$ or NH$_4$OH.

Experimental data was fit to the Langmuir adsorption model, which 
represents homogeneous adsorption and estimates the adsorption maximum
by limiting adsorption to monolayer coverage.
The Langmuir adsorption equation applied for analysis 
is shown in Eq.~\ref{eqk1},\cite{knight1,knight5}
\begin{equation}
q_e=  \frac{ K_L q_{m} [{\rm M}^{n+}]_{\rm eq} }{1+ K_L [{\rm M}^{n+}]_{\rm eq}}
\label{eqk1}
\end{equation}
where 
$q_e$ is the mass normalized equilibrium adsorption of the metal for Mg$^{2+}$
and Cu$^{2+}$ ($\mu$mol/g), $q_m$ is the mass normalized adsorption values 
for these cations ($\mu$mol/g), and $K_{\rm L}$ is the Langmuir constant
($L$/$\mu$mol), The Langmuir fitting parameters were estimated by fitting the
data using the curve fitting function in Igor Pro.  Other experimental
details are given in the S.I.

The experimental results are summarized in Fig.~\ref{fig3} depicting the
results of competitive adsorption of Cu$^{2+}$ and Mg$^{2+}$ on fumed silica
surfaces.  The background electrolyte was 0.01~M NH$_4$NO$_3$ at pH=6.5.
Cu has a higher affinity towards silica surface compared to Mg.  
The adsorption isotherm data is fit with the Langmuir adsorption model in
which the adsorption maximum values, $q_m$, are 14$\pm$3 $\mu$mol/g and
5$\pm$1~$\mu$mol/g, for Cu$^{2+}$ and Mg$^{2+}$, respectively.  Likewise,
the distribution factors and the resulting separation factor of Cu$^{2+}$
over Mg$^{2+}$ maintained values of 4-5 over a large concentration (see the
S.I.).  Although the experimental surface termination is likely different
from the reconstructed $\beta$-cristobalite (001) model used in our AIMD
studies, the measurements and modeling consistently show that Cu$^{2+}$ is
more strongly bound.  A quantitive comparison is currently impossible
because AIMD simulations contain only one Cu$^{2+}$ that cannot form dimers.

Next we address the static distributions in 
the protonation state of the SiOH surface group coordinated to the
cation as desorption proceeds.  Nearest-neighbor SiOH groups are
$\sim$5~\AA\, apart, and H$^+$ exchanges occur via water bridges.
For Na$^+$ (Fig.~\ref{fig4}a), in sampling windows centered around
$Z$$\sim$2.7~\AA\, and below, the silanol group is deprotonated at the
cation binding site.  This maximizes electrostatic attraction with Na$^+$.
The exception is $Z$$\sim1.9$~\AA.  A closer examination reveals
that the Na$^+$ ion is vacillating between two SiO$^-$/SiOH
groups, coordinating to one and then the other.  
When $Z$ is around $\sim$3.0 to~4.0~\AA, $P(Z)$ fluctuates between 0.4 and 1.0.
Asymptotically, when Na$^+$ is infinitely far away, each of the four SiOH on
this surface should have an equal, 25\% probability of being deprotonated.
The AIMD trajectories are not sufficiently long to reflect that.  We stress
that our trajectory lengths are chosen to converge $\Delta W(Z)$ to a certain
precision, not necessarily other properties like mean protonation states.

\begin{figure}
\centerline{\hbox{ \epsfxsize=4.00in \epsfbox{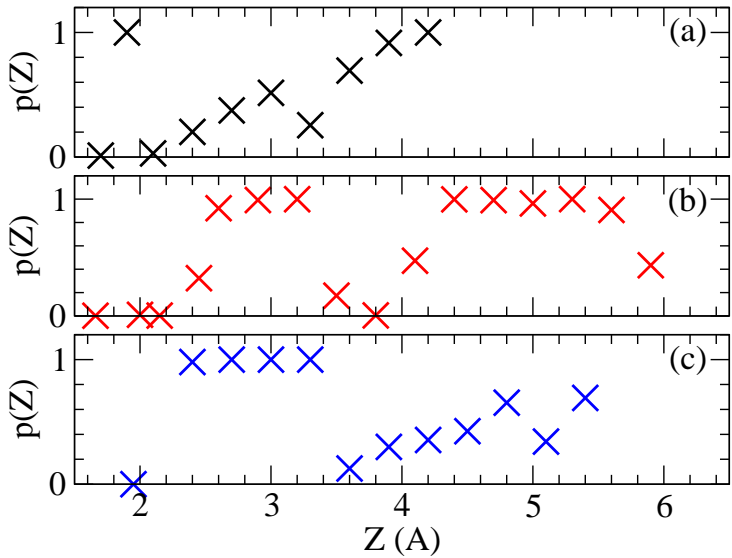} }}
\caption[]
{\label{fig4} \noindent
Protonation states ($p(Z)$) of SiO$^-$ group originally
bonded to M$^{n+}$, as a function of sampling window $Z$.
(a) Na$^+$; (b) Mg$^{2+}$; (c) Cu$^{2+}$.
The SiO$^-$ is ``protonated'' if its O$^-$ ion is
within 1.25~\AA\, of any proton in the simulation cell.
}
\end{figure}

Mg$^{2+}$ and Cu$^{2+}$ exhibit qualitatively similar protonation behavior
(Fig.~\ref{fig4}b,c), despite the fact that their $\Delta W(Z)$ are
significantly different; Cu$^{2+}$ dissociates a H$_2$O at large $Z$
while Mg$^{2+}$ does not; and the error bars in $P(Z)$ may be large.
$P(Z)$$\approx$0 at $Z$$<$2.4~\AA\, presumably because M$^{2+}$ strongly
repels the like-charged H$^+$ in its vicinity.  $P(Z)$$\sim$1 from
2.4~to~3.3~\AA\, the region between inner- and outer-sphere
complexes.  It drops to a low value around 3.6~\AA, then returns to higher
average protonated states.  For both Mg$^{2+}$ and Cu$^{2+}$, $P(Z)$ varies
non-monotonically as the cation desorbs.  This may be a useful
consideration when defining the SiOH protonation as a ``slow variable'' 
in future metadynamics ion desorption calculations.

The SI considers dynamic correlations between SiOH protonation states and
$\Delta W(Z)$.  We conclude that SiOH protonation state fluctuations are
not strongly dynamically correlated with cation desorption free energies
in windows where 0$<P(Z)<$1. This suggests
that AIMD PMF predictions are reliable despite the somewhat
slow $p(t)$ fluctuation time scales.

In conclusion, we have used
{\it ab initio} molecular dynamics potential-of-mean-force techniques to
calculate the desorption free energies
of Na$^+$, Mg$^{2+}$, and Cu$^{2+}$ from a model silica surface with a
4.0/nm$^2$ SiOH surface density and pK$_a$$\sim$7.0-8.1.  The cations are
initially adsorbed in an inner-sphere monodentate manner.  
Na$^+$ at standard state (1.0~M concentration) is predicted to
be unbound, while the divalent cations are predicted to be thermodynamically
stable when coordinated to SiO$^-$ groups by 0.14 and 0.47~eV, respectively,
in inner-sphere configurations.  The predicted trends are in qualitative
agreement with competitive batch adsorption measurements.  We argue that
negatively charged surfaces which allow bi- or tri-dentate SiO$^-$ binding to
metal ions should favor inner-sphere adsorption over outer-sphere adsorption
even more strongly.  Water-bridge-assisted proton exchange between surface
SiOH groups is observed as cation desorption proceeds, which increases
interfacial fluctuations.  The protonation state ($P(Z)$) of the SiO$^-$ group
at the binding site is not a monotonic function of the separation $Z$ between
the cation and the surface.  A water molecular coordinated to Cu$^{2+}$ 
donates a proton to surface SiO$^-$ with predicted pK$_a$ of about
7.0-8.1,\cite{leung2009} forming 4-coordinated Cu$^{2+}$(OH$^-$)(H$_2$O)$_3$
complexes.  To within computational accuracy, this is consistent
with the experimental Cu$^{2+}$(H$_2$O)$_6$ pK$_a$ of about 8.  In contrast,
Mg$^{2+}$ invariably forms octahedral Mg$^{2+}$(H$_2$O)$_6$ or 
Mg$^{2+}$(H$_2$O)$_5$(SiO$^-$) complexes.  
Our conclusion that the Cu$^{2+}$OH$^-$ complexes occur as Cu$^{2+}$ desorbs
also informs speciation choices for future classical-MD investigations.
While AIMD is computationally costly, our predictions will be useful for
benchmarking and assessing the applicability of simpler molecular force fields
for ion-binding application, and paves the way for enumerating desorption
free energies from multiple binding sites in the future.

\noindent{\bf Supporting Information Available:} Details of PMF calculations;
snapshots of cation adsorption on silica surfaces and classical PMF
results conducted using classical force field-based molecular dynamics;
details of batch adsorption measurements; dynamical correlations between
SiOH protonation state and cation desorption

We thank Jessica Rimsza and Jacob Harvey for useful discussions.  
Sandia National Laboratories is a multimission laboratory managed and
operated by National Technology and Engineering Solutions of Sandia, LLC,
a wholly owned subsidiary of Honeywell International, Inc., for the
U.S.~Department of Energy’s National Nuclear Security Administration under
contract DE-NA0003525.   This work is based on materials support
by the U.S. DOE Office of Basic Energy Sciences, Division of Chemical
Sciences, Geosciences, and Biosciences.

\end{document}